\documentclass[a4paper]{AO4ELT}  

\usepackage{microtype}
\usepackage{amsmath,amsfonts,amssymb}
\usepackage{graphicx}
\usepackage{pst-all} 
\usepackage[colorlinks=true, allcolors=blue]{hyperref}
\usepackage{astro_bib_macro_mam}

\makeatletter         
\def\@maketitle{
\includegraphics[width = 170mm]{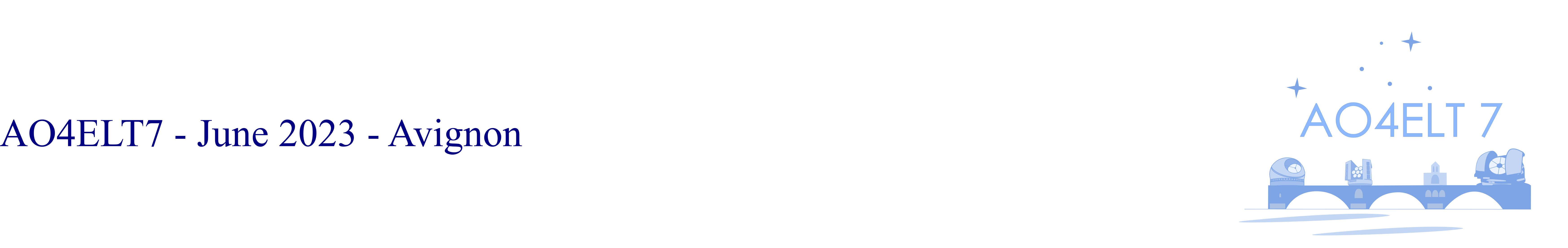}\\[8ex]
\begin{center}
{\Huge \bfseries \sffamily \@title }\\[4ex] 
{\Large  \@author}\\[4ex] 
\@date
\end{center}}

\title{Exoplanet imaging with ELTs: exploring a second-stage AO with a Zernike wavefront sensor on the ESO/GHOST testbed}

\author[a]{M. N'Diaye}
\author[b]{A. Vigan}
\author[c]{B. Engler}
\author[c]{M. Kasper}
\author[c]{S. Leveratto}
\author[b]{J. Floriot}
\author[b]{M. Marcos}
\author[a]{C. Bailet}
\author[b]{K. Dohlen}
\affil[a]{Université Côte d'Azur, Observatoire de la Côte d'Azur, CNRS, Laboratoire Lagrange, France}
\affil[b]{Aix Marseille Université, CNRS, CNES, LAM, Marseille, France}
\affil[c]{European Southern Observatory (ESO), Karl-Schwarzschild-Str. 2, 85748 Garching, Germany}

\authorinfo{Further information, send correspondence to \url{mamadou.ndiaye@oca.eu}}

\begin{document} 
\maketitle
\begin{abstract}
We propose to explore a cascade extreme Adaptive optics (ExAO) approach with a second stage based on a Zernike wavefront sensor (ZWFS) for exoplanet imaging and spectroscopy. Most exoplanet imagers currently use a single-stage ExAO to correct for the effects of atmospheric turbulence and produce high-Strehl images of observed stars in the near-infrared. While such systems enable the observation of warm gaseous companions around nearby stars, adding a second-stage AO enables to push the wavefront correction further and possibly observe colder or smaller planets. This approach is currently investigated in different exoplanet imagers (VLT/SPHERE, Mag-AOX, Subaru/SCExAO) by considering a Pyramid wavefront sensor (PWFS) in the second arm to measure the residual atmospheric turbulence left from the first stage. Since these aberrations are expected to be very small (a few tens of nm in the near-infrared domain), we propose to investigate an alternative approach based on the ZWFS. This sensor is a promising concept with a small capture range to estimate residual wavefront errors thanks to its large sensitivity, simple phase reconstruction and easiness of implementation. In this contribution, we perform preliminary tests on the GHOST testbed at ESO to validate this approach experimentally. Additional experiments with petalling effects are also showed, giving promising wavefront correction results. Finally, we briefly discuss a first comparison between PWFS-based and ZWFS-based second-stage AO to draw preliminary conclusions on the interests of both schemes for exoplanet imaging and spectroscopy with the upgrade of the current exoplanet imagers and the envisioned ExAO instruments for ELTs.
\end{abstract}


\section{INTRODUCTION}\label{sec:intro}
High-contrast imaging and spectroscopy is a fast-evolving field in astronomy to observe circumstellar disks, brown dwarfs and extrasolar planets \cite{Oppenheimer2009,Traub2010,Currie2023,Galicher2023}. It enables the community to retrieve the astro-photometry of these substellar mass companions and analyze the spectral features of their atmosphere. The interpretation of this information gives us clues on the formation, evolution, and diversity of planetary systems. This analysis will ultimately provide insights on the presence of life outside our solar system. 

In the past decade, a new generation of exoplanet spectro-imagers have been deployed on ground-based telescopes to study disks, warm or massive gas giant planets in the vicinity of bright nearby stars \cite{Macintosh2014,Jovanovic2015,Beuzit2019,Males2022}. These high-contrast facilities combine extreme adaptive optics to correct for the effects of the atmospheric turbulence on the image of an observed star \cite{Guyon2018}, coronagraphy to attenuate starlight \cite{Guyon2006,Ruane2018b}, observing strategies and post-processing methods to retrieve the signal of a planetary companion around the star\cite{Cantalloube2022}. The most advanced high-contrast facilities enables to reach a contrast, i.e. the flux ratio between an observed star and its planet, from to $10^4$ to $10^6$ at angular separations down to 200\,mas. Such an angular distance on sky corresponds to a separation of 5$\lambda/D$ for a wavelength $\lambda=1.6\mu$m in the near-infrared ($H$-band) and for a telescope with an aperture diameter $D=$8\,m.

Their ExAO systems typically use a Shack-Hartmann wavefront sensor (WFS) in visible light to measure the wavefront errors due to the atmospheric turbulence, a deformable mirror (DM) to correct for the corresponding aberrations in the image of an observed star in the near infrared, and a real-time computer (RTC) to convert the WFS measurements into applied voltages on the DM for wavefront error correction. The current generation of exoplanet imagers have their ExAO systems running at 1\,kHz, leading to images with Strehl ratios larger than 0.9 in H-band in good observing conditions\cite{Beuzit2019}.  

Different upgrades are currently envisioned for the current facilities \cite{Chilcote2018,Lozi2018,Boccaletti2020,Males2022} to access young gas giant planets down to the snowline, observe a larger number of red stars, and enable a deeper characterization of the planet atmosphere, as suggested in the case of VLT/SPHERE\cite{Boccaletti2020}. To achieve these science goals, there is a need for an increased contrast at a few resolution elements ($\lambda/D$) from an observed star, a gain in sensitivity for the red stars, and enhanced spectroscopic capabilities. In terms of wavefront corrections, these goals possibly translate into an increase in temporal bandwidth of the adaptive optics system and possibly the inclusion of a more sensitive wavefront sensor.  

Several teams are thus investigating an upgrade of the ExAO systems by considering the inclusion of multiple control loops and run systems at a speed faster than 3\,kHz. With the addition of new control loops, these teams often consider the introduction of a pyramid wavefront sensor (PWFS)\cite{Ragazzoni1996} for wavefront measurements. Having two control loops in cascade is an attractive solution to minimize the wavefront error residuals that are currently achieved with the current ExAO systems. A typical emerging solution consists in inserting a second stage AO loop in an existing ExAO instrument with a near-infrared PWFS, a second DM and an additional RTC to run corrections at up to 3\,kHz. Such a solution is currently developed on VLT in the context of SAXO+, the adaptive optics upgrade of the SPHERE instrument\cite{Boccaletti2022,Stadler2022}.   

With the ExAO systems offers highly corrected star images, e.g. with a Strehl ratio larger than 90\% in $H$ band in high flux regime, the wavefront errors are about 85\,nm RMS or 0.05$\lambda$ in H-band, corresponding to the regime for which phase errors are much smaller than 1\,rad. This range is extremely favorable to Zernike wavefront sensors (ZWFS)\cite{Zernike1934} which are known to be very sensitive and accurate in the small aberration regime\cite{Guyon2006,Jensen-Clem2012,N'Diaye2013a,Ruane2020,Chambouleyron2021}. Our idea consists in investigating an alternative second-stage AO loop based on a Zernike wavefront sensor. 

In these proceedings, we recall the principle and formalism of the ZWFS \cite{Zernike1934}. The implementation of the ZWFS-based control loop on the GHOST testbed at ESO is detailed with the main features of the bench. We then present our preliminary results in terms of wavefront error compensation and impact on the contrast on coronagraphic images acquired with a classical Lyot coronagraph (CLC)\cite{Lyot1932,Vilas1987}. Finally, we show some first experiments and results of our control loop for the combination of AO residuals and petalling modes before deriving some first conclusions about the potential of this solution. The contribution is limited to the presentation to the early results of our tests. Further description and analysis of our tests and results will be extensively reviewed in a forthcoming paper (N'Diaye et al. in prep). 

\section{EXPERIMENTAL SETUP AND PROTOCOL}\label{sec:setup}
\subsection{GHOST testbed setup}\label{subsec:setup}
\begin{figure}[!ht]
    \centering
    \includegraphics[scale=0.4]{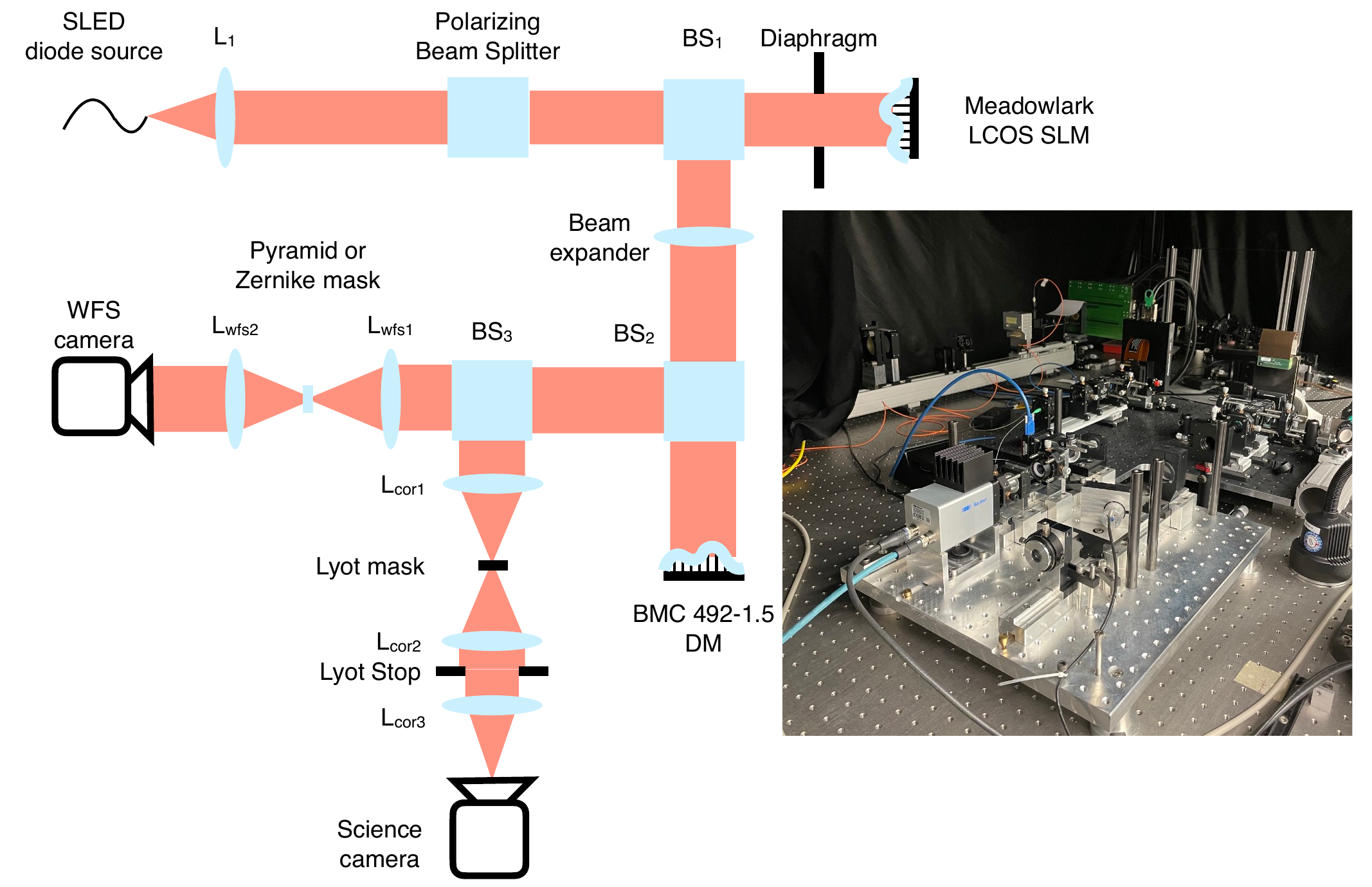}
    \caption{\textbf{Left}: Schematic layout of the GHOST testbed, with the following symbols: L: lens, BS: beam splitter, DM: deformable mirror, SLM: spatial light modulator. See text for more details. The modulation mirror used for the PWFS is not represented in this scheme. \textbf{Right:} Picture of the testbed on March 8$^{th}$, 2023.}
    \label{fig:ghost_layout}
\end{figure}

The GHOST testbed is located at ESO Headquarters in Garching. A scheme of the optical layout and a picture are given in Figure \ref{fig:ghost_layout}. We here briefly recall the main features of the bench with a focus on the parts that are used for our experiments.

GHOST uses a SLED diode source emitting at the wavelength $\lambda$ of 770nm. After being collimated by an achromatic lens L$_1$, the beam goes through a polarizing beam splitter cube and a first standard beam splitter cube (BS$_1$) which sends the light through a 10\,mm diaphragm. The beam then hits the Meadowlark LCOS spatial light modulator (SLM) in reflection to inject wavefront residuals. 

The reflected light goes back to BS1 and goes across a beam reducer before going through a second beam splitter BS$_2$ to send the light to the wavefront corrector arm which includes the Boston micro-machine 492-1.5 DM to compensate for the AO residuals introduced with the SLM. The reflected beam then travels back to BS$_2$. It then reaches a third cube BS$_3$ which equally splits the light between the science arm with a classical Lyot coronagraph (CLC) and the wavefront sensing path with optical parts to measure the AO residuals introduced by the SLM. The CLC includes a 4$\lambda/D$ opaque focal plane mask (FPM) and a Lyot stop with a diameter of 0.84 times the pupil size.  

This wavefront sensing path includes a focusing lens to form the source image on the sensor, a field stop to reduce aliasing effects, the wavefront sensor itself and its 10GigE camera with a f/50 beam ratio to measure wavefront errors. The standard GHOST configuration uses the pyramid wavefront sensor (PWFS) for the wavefront error measurements. An additional PI SL-325 modulation mirror is present to enable the modulated PWFS mode configuration. In this work, we replace the Pyramid with a Zernike mask.  

\subsection{Zernike wavefront sensor}

\subsubsection{Principle}
\begin{figure}[!ht]
    \centering
    \includegraphics[scale=0.67]{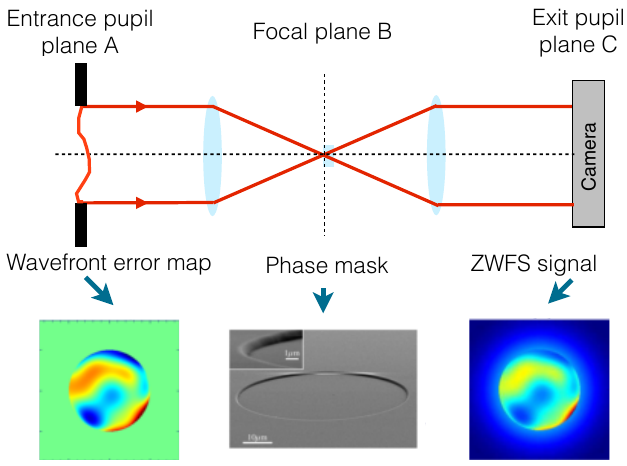}
    \caption{Schematic diagram of the ZWFS analysis with the input pupil-plane wavefront errors to be estimated, a phase-shifting mask centered on the on-axis stellar point source at the focus of the aperture and the re-imaged pupil-plane intensity measurement. For small aberrations ($\varphi \ll 1$rad), a linear reconstruction of the aberrations is performed from the recorded intensity with a nanometric accuracy.}
    \label{fig:zwfs_scheme}
\end{figure}

We briefly recall the principle of the ZWFS \cite{Bloemhof2003,Bloemhof2004,Dohlen2004}, see the optical layout in Figure \ref{fig:zwfs_scheme}. From an entrance pupil with a wavefront error, the system uses a lens to form the image of an observed source in the following focal plane in which a phase-shifting mask with a size of about a resolution element is inserted. This small dot introduces a phase shift at the core of the source point spread function (PSF), leading to intensity variations in the re-imaged pupil plane that are directly related to the entrance pupil aberrations $\varphi$. 

In the small aberration regime ($\varphi \ll 1$\,rad), the intensity variations in the re-imaged pupil are linearly related to the wavefront errors in the entrance pupil, enabling a simple and fast reconstruction of the wavefront errors. The ZWFS principle and formalism has been extensively described in astronomy in the recent years, further details can be found in the literature, e.g. see papers \cite{Jensen-Clem2012,N'Diaye2013a,Ruane2020,Chambouleyron2021}. 

In the past few years, ZWFSs have been the subject of intensive research in astronomy to address different aspects such as non-common path aberrations in high-contrast facilities \cite{N'Diaye2013b,N'Diaye2016b,Vigan2019,Vigan2022,Pourcelot2021,Hours2022}, coarse or fine cophasing of segmented aperture telescopes \cite{Surdej2010,Vigan2011,Janin-Potiron2017,Cheffot2020,vanKooten2022,Salama2022}, low-order wavefront errors \cite{Shi2016,Pourcelot2022,Pourcelot2023}, and picometric precision metrology for future large space observatories with high-contrast capabilities \cite{Ruane2020}. Many innovative ZWFS flavors \cite{Wallace2011,Jackson2016,Doelman2019,Chambouleyron2021}, novel wavefront reconstruction strategies \cite{Moore2018,Steeves2020} have also emerged to increase the sensor capabilities, offering promising potential for extended science return in astronomy. 

More recent works related to ZWFS with adaptive optics towards ELT were presented in this conference, see the respective proceedings. In this paper, we focus on the standard ZWFS and address residual atmospheric wavefront errors that are left after a first ExAO stage. 

\subsubsection{GHOST mask prototype}
THe GHOST ZWFS mask was manufacturing by SILIOS technologies using photolithography with reactive ion etching on a fused silica substrate \cite{N'Diaye2010,N'Diaye2011} to achieve a hole with a 40.5\,$\mu$m diameter and 0.423\,$\mu$m depth.

The prototype was measured at LAM using an interferentiel microscope Wyko NT9100 in vertical scanning interferometry mode, allowing us to confirm that the measured mask diameter and depth are within 1\% of the specifications given to the manufacturer.  

At $\lambda=770$\,nm, with a f/50 beam and a refractive index $n=1.4539$, the mask shows a relative diameter $d=1.05\lambda/D$ and introduced phase shift $\theta=\pi/2$. 

For our experiments, the mask is mounted on a xy-stage mount to ease its alignment with the source image. 

\subsection{Experimental protocol}\label{subsec:zwfs}
In our experiment, we inject maps of residual wavefront errors on the SLM at a frame rate of 422\,Hz. The maps were computed by following the spatial power spectral density (PSD) of the residual wavefront errors after correction of the effects of the atmospheric turbulence based on the VLT/SPHERE characteristics. These maps are calculated for a source magnitude $\Delta$mag=6 and observing conditions with different wind speed and seeing. These phase screens replayed by the SLM at 422Hz come from a simulation with a frame rate (update of atmospheric turbulence) of 2\,kHz where the AO system (SPHERE-like) only runs at 1\,kHz (i.e., 2 turbulence frames are averaged by the WFS and the DM is updated at 1 kHz).

The wavefront sensing and control with ZWFS are performed by first computing an interaction matrix with the recording of the ZWFS response on the camera with 36 pixels across the pupil diameter for the application of push-pull commands of Karhunen–Loève (KL) modes on the DM actuators with 24 actuators across the pupil diameter. We then compute the pseudo-inverse of the interaction matrix to generate the command matrix. 

The temporal control of the wavefront errors relies of the Cosmic RTC \cite{Ferreira2020} which is installed on a GPU server with 2 CPUs and 112 cores in total, and 2 Titan RTX GPUs. It uses a standard AO pipeline which shows a delay of 110$\mu$s between the last pixel received from the WFS and the voltage sent to the DM. The COSMIC Graphics User Interface (GUI) enables to remotely control the DM, the wavefront sensor camera, and the modulation mirror for tests with the PWFS. For the tests with ZWFS control loop, no modulation is required and the modulation mirror was left unused.  

An additional computer with remote access allows us to drive the source illumination, the SLM, the camera in the science arm, and finally the viewer on the wavefront sensing camera with the standalone software. All the functions are remotely accessible independently and an integrated version of all the software is envisioned in the near future.   

Several temporal controllers are also available to explore different control strategies\cite{Nousiainen2022}. In our experiment with the ZWFS, we limit our tests with a classical integrator with gain and leak to mimic a realistic AO control with the second-stage AO. The ZWFS-based control loop runs at the same speed as the SLM (422\,Hz) to simulate a 2nd stage with a speed that is twice the update rate of the first stage. 

The performance with our ZWFS-based control loop is assessed in open and closed loop at the level of the wavefront error residuals and at the level of the measured contrast in the presence of the CLC on the science camera.  

\subsection{Assumptions}\label{subsec:assumptions}
Table \ref{tab:tests_log} gives the different observing conditions of the AO residuals for the different tests which were performed on March 9th, 2023. Additional tests were realized on March 10, 22 ad 24 and the resulting data will be presented in a forthcoming paper. In the following, test 02 represents our baseline in terms of phase residuals on the SLM with the following observing conditions (wind speed of 34m/s, 0.7" seeing) and of the number of corrected modes set to the first 275 KL modes.

\begin{table}[!ht]
    \caption{Summary of the observing conditions used for the tests with the ZWFS-based control loop.}
    \centering
    \begin{tabular}{cccccccccrc}
    \hline\hline
        \textbf{Date} & \textbf{Test} & \textbf{Source} & \textbf{Seeing} & \textbf{Wind} & \textbf{Corrected} & \textbf{Loop} & \textbf{Loop} & \textbf{CLC} & \textbf{DIT} & \textbf{NDIT}\\
         &  & \textbf{flux} &  & \textbf{speed} & \textbf{modes} & \textbf{gain} & \textbf{leak} & \textbf{} & \textbf{} & \textbf{}\\
         & & mA & arcsec & m.s$^{-1}$ & & & & & $\mu$s & \\\hline
         2023-03-09 & 02 & 0.025 & 24 & 0.7 & 275 & 0.5 & 0.99 & YES & 15000 & 2000\\
         2023-03-09 & 02 & 0.025 & 24 & 0.7 & 275 & 0.5 & 0.99 & NO & 100 & 2000\\\hline
    \end{tabular}
    \label{tab:tests_log}
\end{table}

\section{AO RESIDUAL COMPENSATION}\label{sec:compensation}
We here analyze the results of the AO residuals compensation with the ZWFS-based control loop. The observing conditions and the considered parameters correspond to the values considered for Test 02, see Table \ref{tab:tests_log}. We study the evolution of the coefficient corresponding to each KL controlled mode before considering the overall total of residual aberrations by considering the quadratic sum of the coefficients of all the modes. 

\subsection{Open and closed loop performance for single modes}\label{subsec:singlemodes}
\begin{figure}[!ht]
    \centering
    \includegraphics[width=0.67\textwidth]{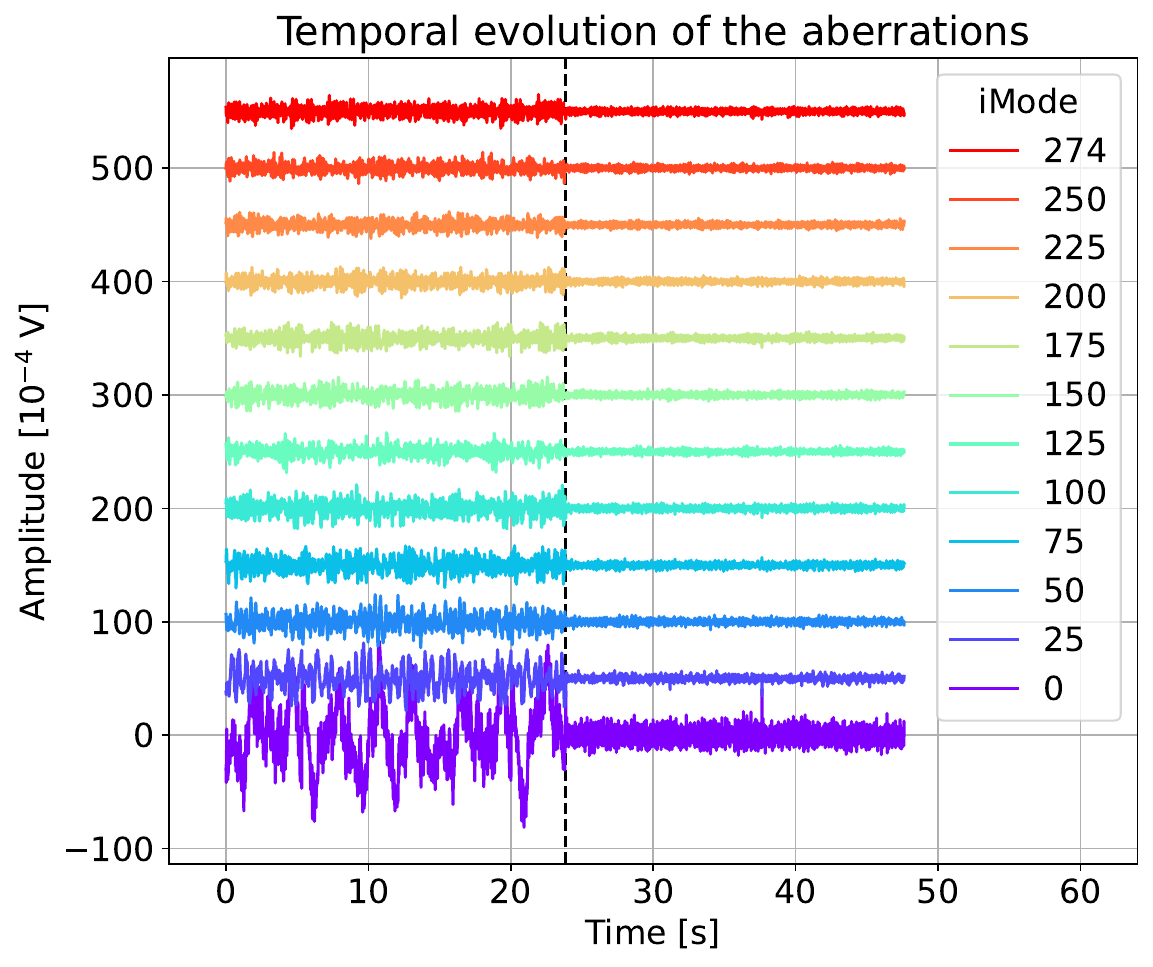}
    \caption{Temporal evolution of the aberrations in open and closed loop with the ZWFS-based control loop (left and right from the vertical dashed line) for some of the considered KL modes. The results are here given under the set conditions for Test 02. While 275 KL modes are controlled with our closed loop, we only display 12 of them for the sake of clarity. For the represented KL modes, the curves have been artificially shifted along the vertical axis to enhance readability. All the curves actually oscillates around zero.}
    \label{fig:singlemodes_timeseries}
\end{figure}

Figure \ref{fig:singlemodes_timeseries} shows the evolution of the coefficients of the KL corrected modes. The ZWFS-based control mode runs in open mode during the first 23.5s and then in closed loop. The dashed vertical line represents the switch from the open to closed loop operation. Qualitatively, we notice a reduction of the standard deviation of the wavefront errors residuals between the open and closed loops for all the represented modes. 

\begin{figure}[!ht]
    \centering
    \includegraphics[width=0.67\textwidth]{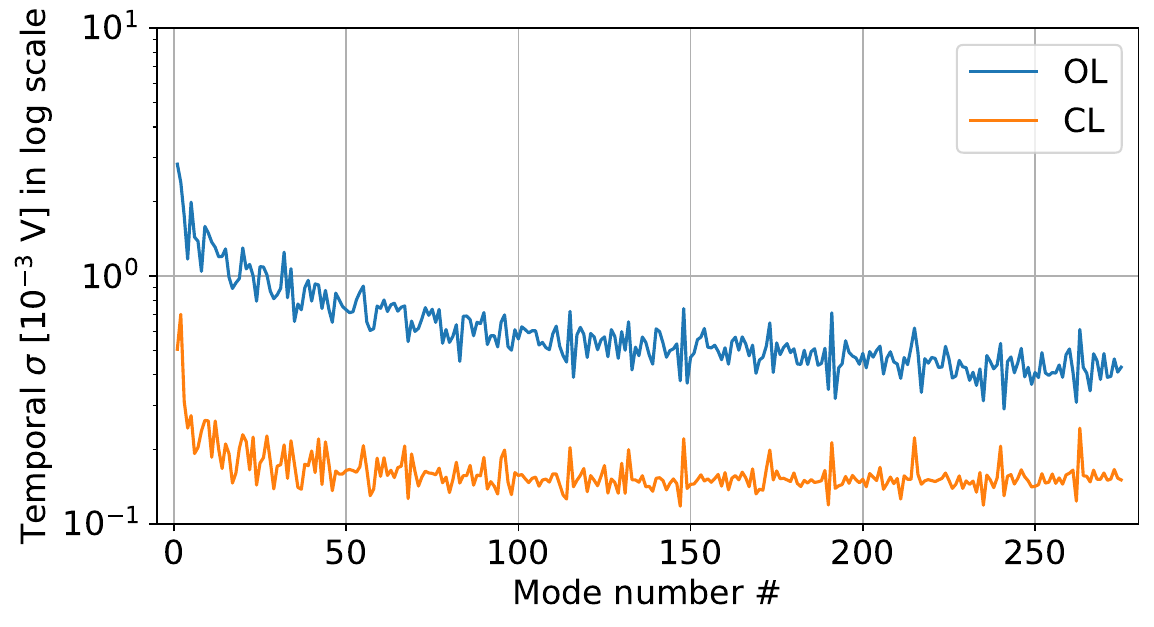}
    \caption{Temporal standard deviation of the KL mode coefficients in open (OL) and closed loop (CL), in log scale. The values are expressed in voltage applied on the DM.  The results are here given under the set conditions for test 02. A clear drop is clearly observed for all the KL mode coefficients from open to closed loop, showing the efficiency of the ZWFS-based control loop. These residual modal coefficients are affected by the non-linearity of the sensor, specially for the OL case for which the values are larger, and so the relative gain should be even larger than what is displayed here.}
    \label{fig:singlemodes_allvalues}
\end{figure}

Figure \ref{fig:singlemodes_allvalues} shows the temporal standard deviation of the wavefront errors in open and closed loops for all the controlled KL modes. Quantitatively, the wavefront errors are reduced by a factor from about 4 to 2 as the mode index number increases, showing a clear improvement of the phase correction for all the modes with the ZWFS-based controlled loop. Since the residual modal coefficients are affected by the non-linearity of the WFS (large values will be damped),  the true residuals are somewhat different to what is displayed in the plot, especially for the open loop case in which the residuals are larger. So, the relative gain may be even a bit larger than 4 to 2. This first result of AO residual compensation with our control loop is encouraging to further reduce the wavefront error residuals in ExAO  systems for exoplanet imaging and spectroscopy.  

\subsection{Open and closed loop performance for all modes}\label{subsec:allmodes}
\begin{figure}[!ht]
    \centering
    \includegraphics[width=0.67\textwidth]{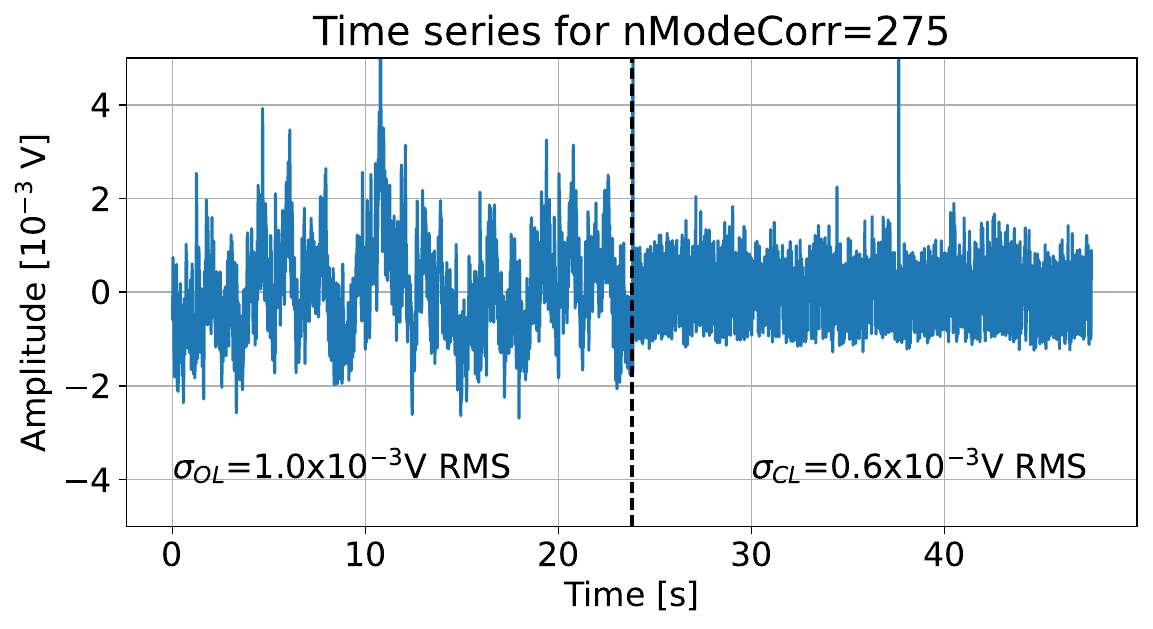}
    \caption{Temporal evolution of the global wavefront errors in open and closed loop with the ZWFS-based control loop (left and right from the vertical dashed line). The values are expressed in voltage applied on the DM. The results are here given under the set conditions for test 02. The temporal standard deviation of the errors decreases by a factor of about 2 from open to closed loop, validating the benefit of our ZWFS-based second stage AO loop on the global wavefront errors.}
    \label{fig:allmodes_timeseries}
\end{figure}

The reduction of AO residuals observed for each KL mode coefficient between open and closed loop is naturally confirmed when we look at the quadratic sum of all the 275 KL controlled modes, see Figure \ref{fig:allmodes_timeseries}. In terms of correction performance, an overall gain of about 2 is observed in AO residuals between open and closed loop configurations.    

\begin{figure}[!ht]
    \centering
    \includegraphics[width=0.67\textwidth]{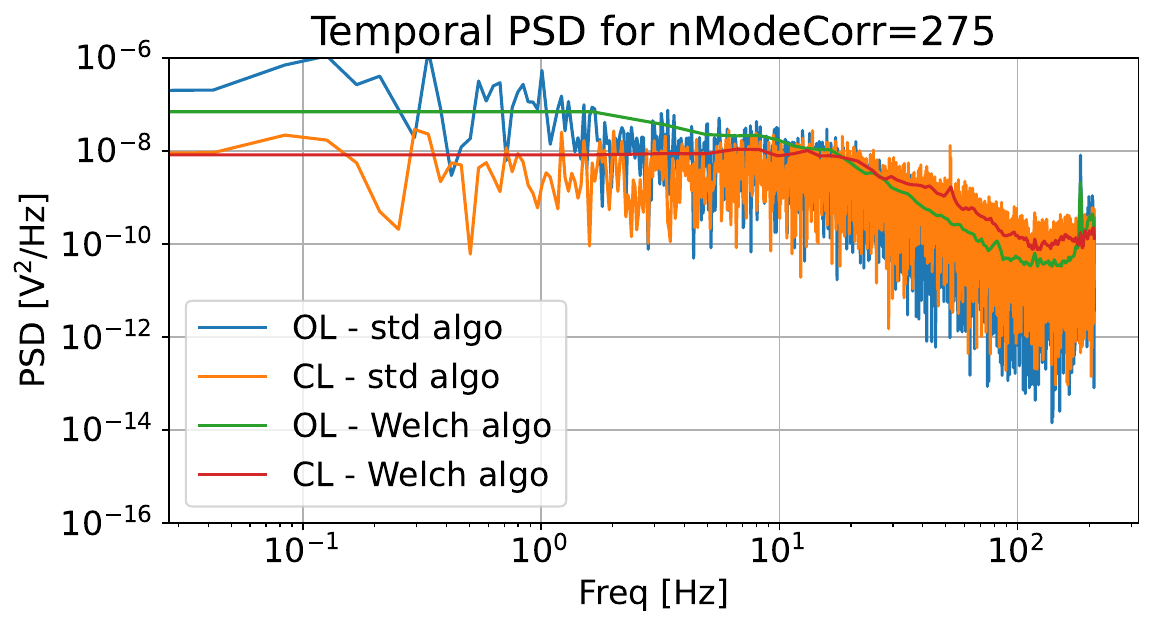}
    \caption{Temporal PSD of the wavefront errors in open (OL) and closed (CL) loops for the results displayed in Figure \ref{fig:allmodes_timeseries} and corresponding to the Test 02. The PSDs are represented in both cases using the standard algorithm and the Welch algorithm. CLosing the ZWFS-based control loop enables us to clearly reduce the temporal PSD amplitude for the temporal frequencies up to 20\,Hz.}
    \label{fig:allmodes_psd}
\end{figure}

We also analyzed the temporal power spectral density (PSD) of the overall AO residuals for the 275 controlled modes, see Figure \ref{fig:allmodes_psd}. As mentioned for Figure \ref{fig:singlemodes_allvalues}, we recall that the measurements are altered by the non -linearity of the sensor, especially in open loop. Between the open and closed loops, the AO residual PSD show a decrease in amplitude for the temporal frequencies up to about 20\,Hz before showing some small performance degradation at larger temporal frequencies. The behavior could be explained by the fact that we worked with a simple integrator as the controller and we did not try to optimize it further. Further analyzes will be made to better understand the origin of this somewhat unexpected behavior. Overall, the temporal PSDs of the AO residuals is reduced with our ZWFS controlled loop, showing good promises to control errors at both different temporal and spatial frequencies. 

\section{IMPACT ON CORONAGRAPHIC PERFORMANCE}\label{sec:coronagraph}
\subsection{Contrast results in open and closed loop}\label{subsec:contrast}
\begin{figure}[!ht]
    \centering
    \includegraphics[width=0.67\textwidth]{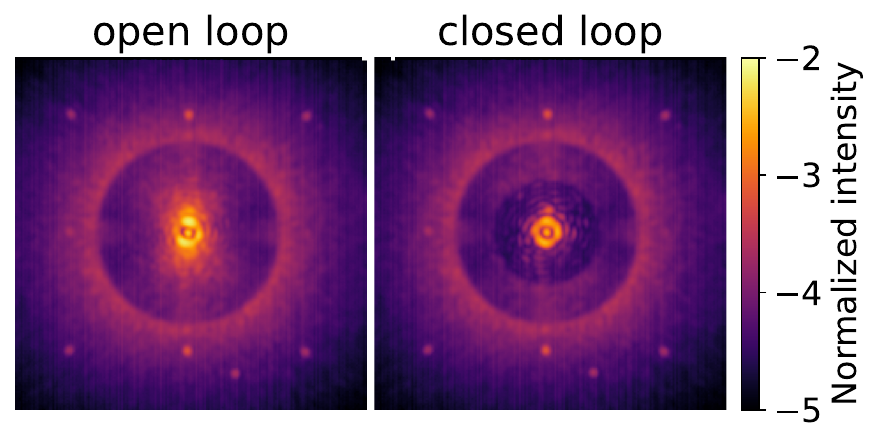}
    \includegraphics[width=0.67\textwidth]{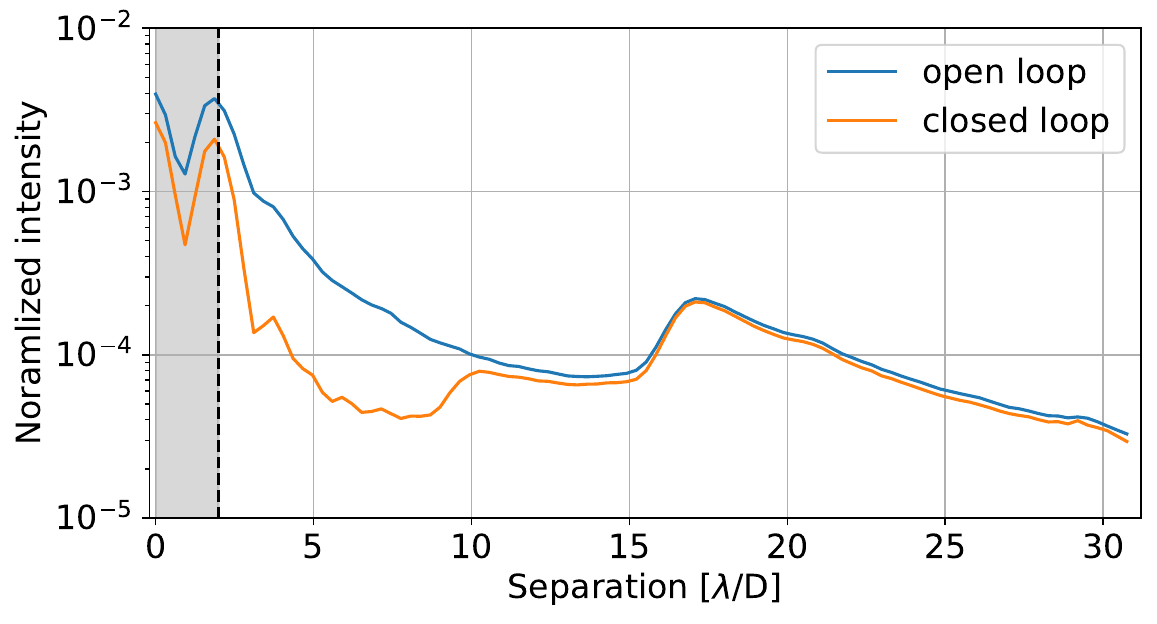}
    \caption{\textbf{Top plot}: Coronagraphic images with a CLC in open and closed loop for the ZWFS-based wavefront control. The coronagraphic images are normalized with respect to the intensity peak of the non coronagraphic images in open and closed loop. In both coronagraphic images, the large bright ring with a radius of about 17$\lambda/D$ delimits the controlled area due the first stage AO loop and symbolized by the AO residuals injected on the SLM. In the coronagraphic image in closed loop, the short bright ring with a radius of about 10$\lambda/D$ denotes the controlled area obtained with the ZWFS-based second stage AO loop, leading to clear contrast improvement at short separations from the source compared to the image in open loop. \textbf{Bottom plot}: Averaged intensity profiles of these coronagraphic images as a function of the angular separation. The shadowed area represents the projection of the coronagraph focal plane mask in the final image plane. A contrast gain up to a factor of about 10 is observed in the coronagraphic profiles from the open to closed loop operation, highlighting the efficiency of the wavefront error correction provided by our ZWFS-based control loop on the science image. The residual intensity is most likely dominated by the noncommon path aberrations that are currently left uncorrected in the testbed.}
    \label{fig:corona}
\end{figure}

We here study to impact of the AO residual compensation on the science image. In the following, we compare the residual intensity of the source image with the CLC on the science camera in open and closed loop. 
Figure \ref{fig:corona} top plot shows the coronagraphic images with CLC in open and closed loops. The images have been normalized with respect to the intensity peak of the images in the absence of CLC in open and closed loops. 

Both images exhibit a common large dark zone which is related to the AO residuals injected on SLM and controlled up to a given spatial frequency of 15 cycles/pupil [cyc/p]. More interestingly, a second and smaller dark zone is observed on top of the large dark zone for the closed-loop images, showing a contrast improvement at the shortest separations from the source. This effect clearly results from the AO residual compensation with the ZWFS-based control loop. 

Figure \ref{fig:corona} bottom plot represents the averaged intensity profiles of the coronagraphic images in open and closed loop. A contrast gain of 5 to 10 is observed at the shortest separation from the source, showing good promises for the observation of fainter substellar mass companions at these distances. The residual contrast with the ZWFS loop closed is mostly likely dominated by the quasi-static features existing between the wavefront sensing arm and the science path in the testbed. These somewhat non-common path errors are currently left uncorrected in the testbed. With a standard CLC, we show that our second stage AO loop enables to produce images with deeper contrast at the shortest separations from an observed source. This first encouraging result represents a first in-lab demonstration of the ZWFS-based second stage AO for exoplanet imaging and spectroscopy.

\section{PRELIMINARY TESTS WITH PETALLING EFFETCS}
\subsection{Compensation of the petalling effects}
Petalling effects or low-wind effects are observed in some of the current AO facilities \cite{Sauvage2016,Milli2018,N'Diaye2018,Cantalloube2019,Bos2020,Bertrou-Cantou2022} and expected on future large ground-based observatories. Such effect will degrade the contrast provided by coronagraphic devices, preventing the observation of the faintest planets and circumstellar disks around a nearby star. Compensating this effect is crucial to increase the science return of current and future AO facilities on ELTs with degraded observing conditions. We propose to investigate the efficiency of the ZWFS-based control loop in the presence of AO residuals and petalling effects.

For the tests, we consider a centrally obscured circular aperture with spider struts which split the pupil into 6 petals. AO residuals are injected on the SLM following the observing conditions described in Table \ref{tab:tests_log} and oscillating petalling modes are added with a frequency of 4\,Hz. We build an interaction matrix based on the response of the ZWFS to KL modes and petalling modes in push-pull commands and from the pseudo-inverse of the resulting matrix, we derive the command matrix to control the AO residual modes and the petalling modes. For our tests, the control is limited to the first 100 modes, enabling to compensate for both petalling modes and AO residuals. 

\begin{figure}
    \centering
    \includegraphics[width=0.67\textwidth]{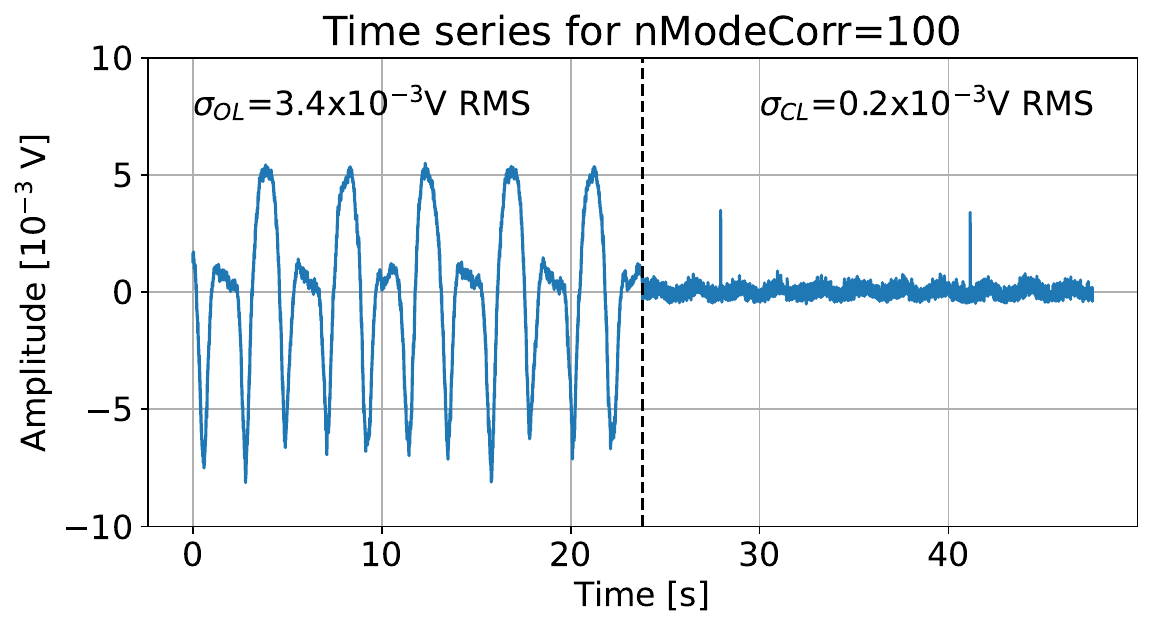}
    \caption{Temporal evolution of the global wavefront errors in open and closed loop with the ZWFS-based control loop (left and right from the vertical dashed line) in the presence of  petalling modes oscillating at 4\,Hz on top of the AO residuals injected on the SLM. The values are expressed in voltage applied on the DM. The results are here given under the set conditions for test 17. The temporal standard deviation of the errors decreases by a factor of about 17 from open to closed loop, validating the efficiency of our ZWFS-based second stage AO loop in the presence of petalling effects.}
    \label{fig:petalling-temporalseries}
\end{figure}

Figure \ref{fig:petalling-temporalseries} shows the temporal evolution of the wavefront errors in open and closed loop for our ZWFS-based second stage AO loop. Oscillating petalling modes with large amplitude have been injected to the SLM, leading to an overall wavefront errors with AO residuals which are about three times larger than the wavefront errors in our tests with standalone AO residuals showed in Figure \ref{fig:allmodes_timeseries}. The temporal standard deviation of the RMS wavefront errors was reduced by a factor of about 17 from open to closed loop, underlining a clear improvement provided by our second stage. More interestingly, the amplitude of the oscillations of the wavefront errors has been strongly attenuated, confirming the ability of our ZWFS-based control loop to control the errors due to the petalling modes. 

\begin{figure}
    \centering
    \includegraphics[width=0.67\textwidth]{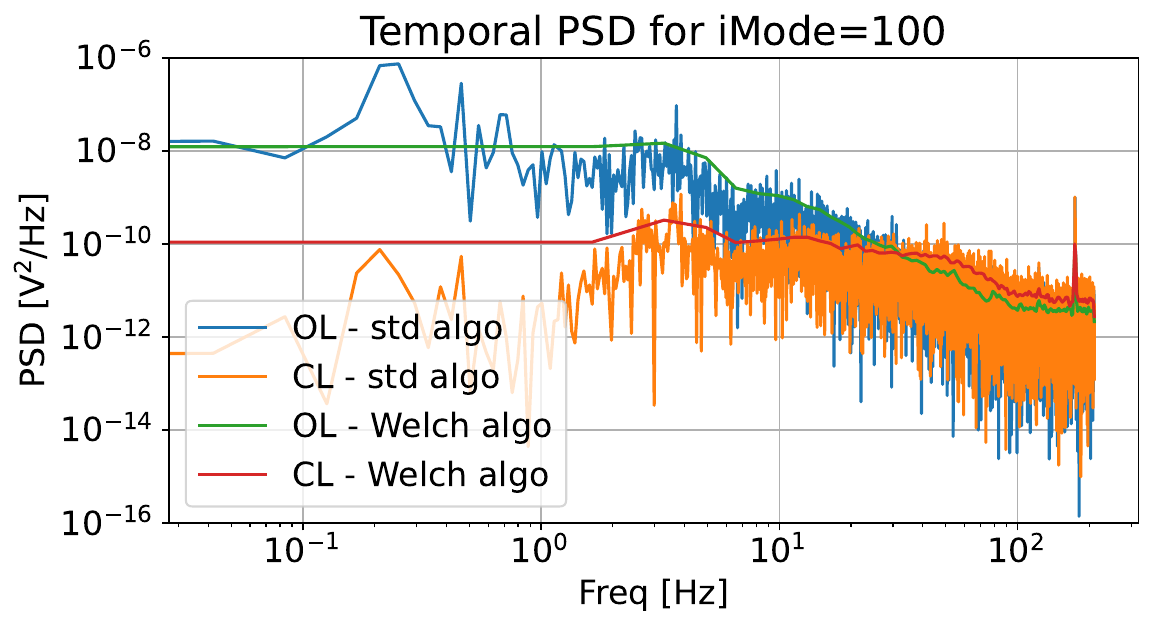}
    \caption{Temporal PSD of the wavefront errors in open (OL) and closed (CL) loops for the results displayed in Figure \ref{fig:petalling-temporalseries} and corresponding to the test 17, in the presence of oscillating petalling modes on top of the AO residuals injected on the SLM. The PSDs are represented in both cases using the standard algorithm and the Welch algorithm. Closing the ZWFS-based control loop enables us to clearly reduce the temporal PSD amplitude for the temporal frequencies up to 30\,Hz and the bump at 4Hz due to the oscillating petalling modes.}
    \label{fig:petalling-temporalpsd}
\end{figure}

This result is confirmed by the observation of the temporal PSD of the wavefront errors, see Figure \ref{fig:petalling-temporalpsd}. Our ZWFS-based control loop reduces the amplitude of the PSD for the temporal frequencies up to 30\,Hz. The bump observed at 4\,Hz in open loop due to the periodicity of the injected petalling modes is clearly reduced by about a factor of 5-10 in closed loop mode. This preliminary result represents a first validation of the ability of our control loop on the control of a combination of AO residuals and petalling modes. Further tests will explore the parameter space of observing conditions and mode control to determine the efficiency of this second-stage AO loop for current and future ground-based observatories.    

\subsection{Impact of the coronagraphic image}
\begin{figure}
    \centering
    \includegraphics[width=0.67\textwidth]{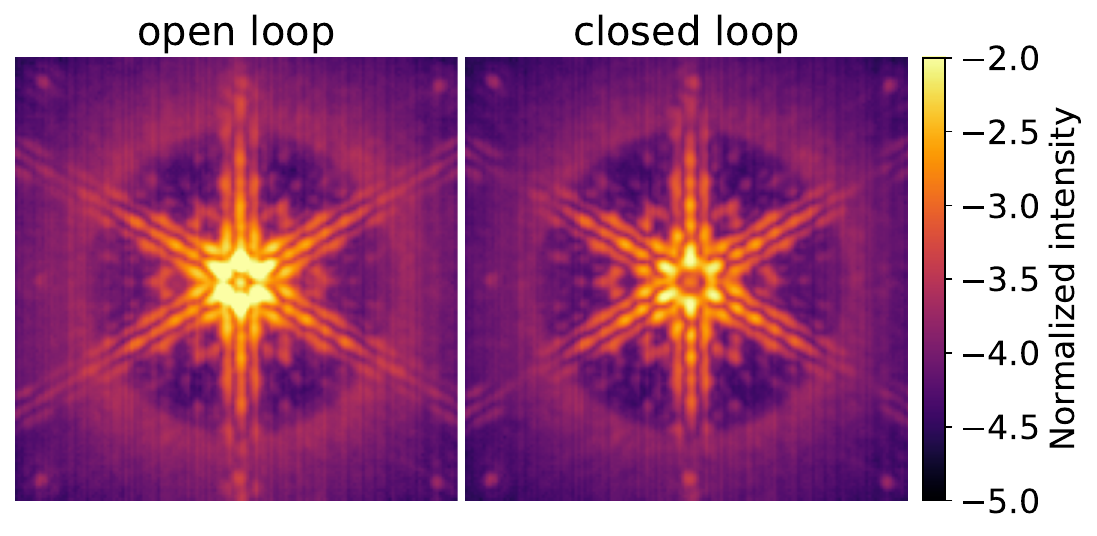}
    \includegraphics[width=0.67\textwidth]{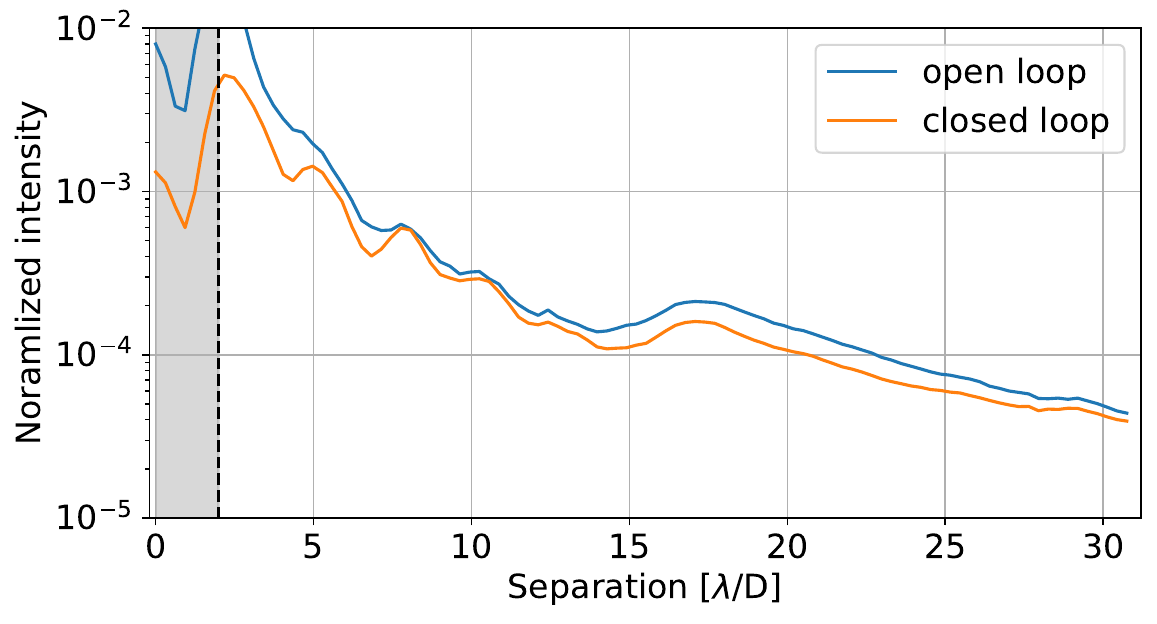}
    \caption{\textbf{Top plot}: Coronagraphic images with a CLC in open and closed loop for the ZWFS-based wavefront control in the presence of petalling modes oscillating at 4\,Hz on top of the AO residuals injected on the SLM. The coronagraphic images are normalized with respect to the intensity peak of the non coronagraphic images in open and closed loop. In both coronagraphic images, the large bright ring with a radius of about 17$\lambda/D$ delimits the controlled area due the first stage AO loop and symbolized by the AO residuals injected on the SLM. The contrast here is clearly limited by the CLC which is not optimized for the ELT pupil (no spiders and central obscuration in the Lyot stop). The effect of the ZWFS-based controlled loop is observable at the shortest separations; the coronagraphic image in closed loop is fainter than its homologous in open loop. \textbf{Bottom plot}: Averaged intensity profiles of these coronagraphic images as a function of the angular separation. The shadowed area represents the projection of the coronagraph focal plane mask in the final image plane. An attenuation of the core image is observed in the coronagraphic profiles at separation shorter than 5$\lambda/D$ from the open to closed loop operation, showing the efficiency of the wavefront error correction provided by our ZWFS-based control loop in the presence of petalling effects on the science image.}
    \label{fig:corona-petalling}
\end{figure}

We also observe the impact of the wavefront correction provided by our ZWFS-based control loop on the contrast in science images. With a Lyot stop as a clear aperture (no central obscuration nor spiders), the CLC configuration is sub-optimal for the petalled pupil we have considered here. Figure \ref{fig:corona-petalling} top plot shows the coronagraphic images before and after closing the loop in the presence of a combination of AO residuals and oscillating petalling modes. The overall contrast is clearly dominated by the diffraction effects due to the pupil spiders. The central part of the coronagraphic image is clearly attenuated with the presence of our control loop. Figure \ref{fig:corona-petalling} bottom plot shows the corresponding normalized intensity profiles of these images, underlining the increase in contrast by a factor of up to ten at angular separations shorter than 5$\lambda/D$. 

This reduction is particularly relevant in the context of the imaging and spectroscopy of faint companions at short orbit of their host stars. Our first tests with our controlled loop confirm our ability to enhance the coronagraphic image quality at short separations in the presence of petalling modes. In the near future, we will investigate the impact of our control loop for different observing conditions, different number of controlled modes and the use of different controllers to maximize the efficiency of our ZWFS-based control loop. 

\section{CONCLUSION AND PROSPECTS}\label{sec:conclusion}
In the contribution, we have introduced a cascaded ExAO system with a second-stage loop based on ZWFS measurements to correct for the first-stage AO residuals for exoplanet imaging and spectroscopy. The ZWFS-based control loop is a simple and suitable solution for cascade AO in the regime of small AO residuals. Its linear phase reconstructor enables the implementation of a second AO loop with one of the most sensitive wavefront sensors at a speed of several times faster than the first AO loop. 

With experiments on the GHOST testbed, we perform a first proof of concept of this approach, showing a reduction by a factor of 2 of the atmospheric wavefront errors left by the first stage AO and a contrast gain up to a factor of 10 at a few $\lambda/D$ separation. These results were achieved for a second-stage AO loop running twice faster than the first stage AO loop. 

We here presented the results for a specific set of observing conditions, source flux, and number of controlled modes. Further tests will explore the parameter space to derive the efficiency of the loop in different conditions and derive the best range for optimal use of our ZWFS-based second-stage AO loop on the current and future exoplanet imagers. 

First tests have also been performed with PWFS and ZWFS on GHOST testbed to compare their performance. Our preliminary results shows very similar results in terms of wavefront corrections and contrast enhancement. A more detailed study of the comparison for different observing conditions, source flux and control parameters will be presented in a forthcoming paper (N'Diaye et al. in prep.). Such a study will enable to determine the best functioning points for both second-stage AO loops and emphasize the possible complementarity between both approaches. 

In our current experiments, we have only considered a simple integrator as a  controller to drive our approach. More advanced controllers will further be investigated with different predictive control strategies (iterative, data-driven machine learning)\cite{Nousiainen2022}. 

As our tests have been performed in monochromatic light, further studies are required to demonstrate the capability of our ZWFS-based control loop in broadband light and exploit photons in wide spectral bandpass. We will address this point in our forthcoming studies to achieve an optimal use of photons in broadband light for exoplanet observations with a two-stage ExAO. 

While our experimental results were achieved in laboratory, on-sky tests are foreseen by Cissé et al. (this proc.) with a dual ZWFS on the PAPYRUS testbed \cite{Muslimov2021} in the near future.

Due to its simplicity and ease of implementation, a ZWFS-based control loop represents an attractive solution towards high-contrast imagers with ELT and could be envisioned as a second option for VLT/SPHERE+ with the implementation of a filter wheel including both a PWFS and a ZWFS. 

\acknowledgments 
 
This work was supported by the Action Spcifique Haute Résolution Angulaire (ASHRA) of CNRS/INSU co-funded by CNES. M. N'Diaye acknowledges support from Observatoire de la Côte d'Azur and Laboratoire Lagrange through the 2022 BQR OCA and 2023 BQR Lagrange programs for the manufacturing of Zernike phase masks and the missions to ESO Garching. AV acknowledges funding from the European Research Council (ERC) under the European Union's Horizon 2020 research and innovation programme, grant agreement No. 757561 (HiRISE).


\bibliography{2023_mndiaye_ao4elt7} 
\bibliographystyle{AO4ELT} 

\end{document}